\documentclass[12pt,superscriptaddress]{article}
\usepackage{palatino,cite,epsfig,amsmath,amssymb}

\allowdisplaybreaks

\begin{document}

\begin{flushright}
WUE-ITP-2004-004\\[-0.15cm] 
IFT-06/04\\[-0.15cm]
%hep-ph/0212169\\[-0.15cm]
January 2004
\end{flushright} 

\begin{center}

{\Large\bf Supersymmetric Lepton Flavour Violation\\ 
at the LHC and LC }

\vspace{0.4cm}

%{\Large\bf LHC / LC Study Group Working Document%
%\footnote{
%For further informations, see {\tt
%http://www.ippp.dur.ac.uk/$\sim$georg/lhclc/ }. For questions and comments,
%please contact {\tt Georg.Weiglein@durham.ac.uk }.}}
 
\vspace{1cm}

{\sc 
F. Deppisch$^{1}$,
J. Kalinowski$^{2}$,
H. P\"as$^1$,
A. Redelbach$^1$,
R. R\"uckl$^1$
}

\vspace*{1cm}

{\sl
$^1$
Institut f\"ur Theoretische Physik und Astrophysik,
Universit\"at W\"urzburg, D-97074 W\"urzburg, Germany

\vspace*{0.4cm}

$^2$
Institute of Theoretical Physics, Warsaw University, Warsaw, Poland}

%\vspace*{0.4cm}

%$^3$Institute 3

\end{center}

\vspace*{1cm}

\begin{abstract}
In supersymmetric extensions of the Standard Model, 
the Yukawa and/or mass terms of the heavy neutrinos can generate
lepton flavour violating slepton mass terms. These new
supersymmetric 
sources of lepton flavour violation may both enhance the rates of 
charged
lepton flavour violating processes, 
$l_\alpha \to l_\beta \gamma$, and   
generate   
distinct final states, like $l_\beta l_\alpha + {\rm jets} + {E\!\!\!/}_T$, 
at future colliders. 
First, we discuss the sensitivity of future $e^+e^-$ colliders 
to the SLFV independently of the lepton flavour violating mechanism.  
Second, we study lepton flavour violating slepton pair production and decay 
at a future $e^+e^-$ linear collider in the context of the 
seesaw mechanism in mSUGRA post-LEP benchmark scenarios. 
We investigate the correlations of these signals with the corresponding 
lepton flavour violating rare decays $l_{\alpha} \rightarrow l_{\beta} 
\gamma$, and show that these correlations 
are particularly suited for probing the origin of lepton flavour violation.
%
%
%We discuss the sensitivity of the LHC and LC to lepton flavour violation in 
%supersymmetric models. We focus on the
%complementarity of high-energy collision processes to the study of the
%rare decays $l_i \to l_j \gamma$. After discussing a simulation of the
%interesting signals independently of the mechanism generating lepton flavour 
%violation, a case study of the lepton flavour violation generated in the 
%seesaw mechanism is presented.
\end{abstract}

\newpage
\section{Introduction}
Neutrino oscillations imply the violation of individual lepton flavours
and raise the interesting possibility of observing 
lepton flavour violation in processes with charged leptons, such
as $\mu\rightarrow e\gamma$ or $\tau\rightarrow\mu\gamma$. 
In the Standard Model these processes are
strongly suppressed due to small neutrino masses. In the
supersymmetric extension of the Standard Model, however,  
the situation may be quite different. 
For example,  the slepton mass matrices need
not simultaneously be diagonalized with the lepton mass matrices.  
When 
sleptons are rotated to the mass eigenstate basis, the slepton mass
diagonalization matrices $W_{i\alpha}$ enter the chargino and neutralino
couplings
\begin{eqnarray}
\tilde{e}_{i} (W^*_{\tilde{l}})_{i\alpha} \bar{e}_{\alpha} \tilde{\chi}^0 
+\tilde{\nu}_{i} (W^*_{\tilde{\nu}})_{i\alpha} \bar{e}_{\alpha} \tilde{\chi}^- 
+\ldots 
\end{eqnarray}
and mix lepton flavour (Latin and Greek subscripts 
refer to the mass-eigenstate and flavour 
basis, respectively). 
Contributions from virtual slepton exchanges can therefore
enhance the rates of rare decays like $\mu\to e\gamma$. 
Furthermore, once superpartners are discovered, the supersymmetric lepton
flavour violation (SLFV) can also be searched for directly at future
colliders where the signal will come  from the production
of real sleptons (either directly or from chain decays of other
sparticles), followed by their subsequent decays. 
Searches for SLFV at colliders have a number of
advantages: superpartners can be 
produced with large cross-sections, flavour violation in the  production and 
decay
of sleptons occurs at tree level 
and therefore
is suppresed only by powers of  $\Delta
m_{\tilde{l}}/\Gamma_{\tilde{l}}$ \cite{feng} in contrast to the
$\Delta m_{\tilde{l}}/m_{\tilde{l}}$ suppression in radiative lepton decays,
where SLFV occurs at one-loop (\cite{Deppisch:2002vz} and references therein). 
Generally, respecting
the present bounds on rare
lepton decays, large SLFV signals are possible both at the LHC
\cite{LHC} and at $e^+e^-$ colliders
\cite{feng,ACFH205,nojiri,GKR,PM,Deppisch:2003wt}. This suggests that    
in some cases the LHC and future $e^+e^-$ colliders may provide 
competitive tools to search for and
explore supersymmetric lepton flavour violation.

In this note we first discuss the sensitivity at future $e^+e^-$ colliders 
to SLFV independently of the lepton flavour violating mechanism.  
The simulation has been performed assuming
a simplified situation with a pure 2-3 intergeneration mixing
between $\tilde\nu_\mu$ and $\tilde\nu_\tau$, and ignoring any mixings
with $\tilde\nu_e$. In the analysis 
the mixing angle $\tilde{\theta}_{23}$ and $\Delta \tilde{m}_{23} =
|m_{\tilde\nu_2} - m_{\tilde\nu_3}|$ have been taken as free, independent
parameters \cite{GKR}. 

In the second part, SLFV generated by the seesaw mechanism is
considered. The heavy right-handed Majorana neutrinos give rise not
only to light neutrino masses but also to mixing of different slepton
flavours due to the effects of the heavy neutrinos
on the renormalization-group running of
the slepton masses.  The implications of recent neutrino measurements
on this mixing are investigated. Moreover we  emphasize the
complementarity of the radiative decays $l_\alpha \rightarrow
l_\beta \gamma$
%in the SUSY seesaw model \cite{Deppisch:2002vz} 
and  the 
%%% INSERT 
specific 
%%% END
lepton flavour violating processes 
$e^{\pm}e^-\rightarrow l_{\beta}^{\pm} 
l_{\alpha}^- \tilde{\chi}_b^0 \tilde{\chi}_a^0$ 
involving slepton pair production and subsequent decay  
\cite{Deppisch:2003wt}.

\section{Sensitivity at future $e^+e^-$ colliders to SLFV
\label{sfcs}}

In discussing the SLFV collider signals at future colliders, one 
has to distinguish two cases in which an
oscillation of lepton flavour can occur: in processes with 
slepton pair production and in processes with 
single slepton production, which differ in the interference
of the intermediate sleptons \cite{feng}. 
Slepton pair production is the dominant 
mechanism at lepton colliders, but it may also occur at hadron
colliders via the Drell-Yan process.
Single  sleptons  may be produced in
cascade decays of heavier non-leptonic superparticles. Such processes are
particularly important for
hadron colliders,  but they may also be relevant for lepton colliders where
a single slepton can be the decay product of a chargino or neutralino. 

The amplitudes for pair production,   
$ \bar f\,f\to \tilde{l}^+_i \, \tilde{l}^-_i
\to l^+_\alpha \, X\, l^-_\beta\, Y $, 
and single production,  $f\,f'\to l^+_\alpha \, X\,  \tilde{l}^-_i
\to l^+_\alpha \, X\, l^-_\beta\, Y$, read, e.g., 
\begin{eqnarray}
&&{\cal M}^{\rm pair}_{\alpha\beta}= \sum_i {\cal M}^{\rm pair}_P  \frac{i}
{q^2-\tilde{m}^2_i+i\tilde{m}_i\Gamma_i} W_{i\alpha} {\cal M}^+_D  \frac{i}
{p^2-\tilde{m}^2_i+i\tilde{m}_i\Gamma_i} W^*_{i\beta} {\cal M}^-_D 
\mbox{~~~ (s-channel)~~~}
\label{amplseins}
\\
&&{\cal M}^{\rm sin}_{\alpha\beta}= \sum_i {\cal M}^{\rm sin}_P W_{i\alpha} \frac{i}
{q^2-\tilde{m}^2_i+i\tilde{m}_i\Gamma_i} W^*_{i\beta} {\cal M}^-_D
\label{amplszwo}
\end{eqnarray}
where ${\cal M}_P$ and ${\cal M}_D$ are the respective production   
and decay
amplitudes for sleptons in the absence of SLFV, and 
$W_{i\alpha}$ stands for the lepton flavour mixing matrix element.

For nearly degenerate
in mass and narrow sleptons, $\Delta \tilde{m}_{ij} \ll \tilde{m}$
and
$\tilde{m}\overline{\Gamma}_{ij} 
\simeq (\tilde{m}_i\Gamma_i+\tilde{m}_j\Gamma_j)/2\ll
\tilde{m}^2$,  the products of slepton 
propagators  can be simplified as follows
\begin{eqnarray}
\frac{i}{q^2-\tilde{m}^2_i+i\tilde{m}_i\Gamma_i}\;\frac{-i}{q^2- 
\tilde{m}^2_j-i\tilde{m}_j\Gamma_j} 
 \sim  \frac{1}{1+i\, \Delta \tilde{m}_{ij}/ \overline{\Gamma}_{ij}}\;
 \frac{\pi}{\tilde{m}\overline{\Gamma}_{ij}} 
\; \delta(q^2-\tilde{m}^2).
\label{propags}
\end{eqnarray}
Then, in the case of 
2-3 intergeneration  mixing, the cross-sections for the above processes
(\ref{amplseins}, \ref{amplszwo}), 
take a particularly simple form \cite{jk01}:     
\begin{eqnarray}
&&\sigma^{\rm pair}_{\alpha\beta}=\chi_{23}(3-4 \chi_{23})  
\sin^2 2\tilde{\theta}_{23} \;
\sigma(\bar f\,f\to \tilde{l}^+_\alpha \,
\tilde{l}^-_\alpha) 
Br(\tilde{l}^+_\alpha \to l^+_\alpha \, X)  
Br(\tilde{l}^-_\alpha \to  l^-_\alpha\, Y) \label{crosspair}\\[2mm]
&&\sigma^{\rm sin}_{\alpha\beta}=\chi_{23} \sin^2 2\tilde{\theta}_{23}\;
\sigma(f\,f'\to l^+_\alpha \, X\, \tilde{l}^-_\alpha)
Br( \tilde{l}^-_\alpha \to l^-_\alpha\, Y) 
\end{eqnarray}
where  $\sigma(f\,f'\to l^+_\alpha \, X\, \tilde{l}^-_\alpha)$, 
$\sigma(\bar f\,f\to \tilde{l}^+_\alpha \,
\tilde{l}^-_\alpha)$ and $Br( \tilde{l}^\pm_\alpha \to l^\pm_\alpha X)$ 
are the corresponding  cross-sections 
and branching ratios in the absence of flavour violation.
The slepton flavour violating mixing effects are encoded in   
\begin{eqnarray}
\chi_{23} = \frac{x_{23}^2}{2(1+x_{23}^2)} \qquad {\mbox {\rm and}} \qquad 
\sin^2 2\tilde{\theta}_{23}
\end{eqnarray}
where $x_{23} = \Delta \tilde{m}_{23}/\overline{\Gamma}_{23}$. 
In the limit $x_{23}\gg 1$, $\chi_{23} $ approaches 1/2, the
interference can be neglected and the cross-sections behave as $\sigma
\sim \sin^2 2\tilde{\theta}_{23}$ . In the opposite case, the interference
suppresses the flavour changing processes, and $\sigma \sim (\Delta
\tilde{m}_{23} 
\sin2\tilde{\theta}_{23})^2$.

To assess the sensitivity of a 500 GeV  $e^+e^-$ linear collider 
to the SLFV, the following processes have been analysed
\begin{eqnarray}
e^+e^- & \rightarrow &
\tilde{\nu}_i\tilde{\nu}^c_j  \rightarrow  \tau^\pm\mu^\mp \tilde{\chi}^+_1
\tilde{\chi}^-_1 \label{snulfv}\\
e^+e^- & \rightarrow &
\tilde\chi^+_2\tilde{\chi}^-_1   \rightarrow  \tau^\pm\mu^\mp 
 \tilde\chi^+_1\tilde\chi^-_1  \label{charlfv}\\
e^+e^- & \rightarrow &
\tilde\chi^0_2\tilde{\chi}^0_1   \rightarrow  \tau^\pm\mu^\mp 
 \tilde\chi^0_1\tilde\chi^0_1  \label{neutlfv}
\end{eqnarray} 
Here $\tilde{\chi}^\pm_1 \rightarrow \tilde{\chi}^0_1 f\bar{f}'$, and
$\tilde{\chi}^0_1$ escapes detection.  The signature of SLFV  would
be $\tau^{\pm}\mu^{\mp}+
\mbox{4 jets}+ {E\!\!\!/}_T$,  $\tau^{\pm}\mu^{\mp}+ 
\ell + \mbox{2 jets}+ {E\!\!\!/}_T$, or $\tau^{\pm}\mu^{\mp}+ {E\!\!\!/}_T$, 
depending on the hadronic or leptonic
$\tilde{\chi}^\pm_1$ decay mode. The purely leptonic decay modes are
overwhelmed by background. In particular, the neutralino pair
production process (\ref{neutlfv}), 
which could still be open if the second chargino and sleptons were too
heavy for (\ref{snulfv}) and (\ref{charlfv}), 
is difficult to extract from background. On
the other hand, with charginos decaying hadronically, the signal 
$\tau^{\pm}\mu^{\mp}+ 4\mbox{ jets}+ {E\!\!\!/}_T$ comes from both 
processes (\ref{snulfv}) and (\ref{charlfv})    
and is SM-background free. The flavour-conserving processes analogous
to (\ref{snulfv}) and (\ref{charlfv}), but with two $\tau$'s in the final
state where  one of the $\tau$'s
decays leptonically to $\mu$, contribute to the
background. On the other hand, if 
jets are allowed to overlap,    
an important SM background to the final states with $\tau^\pm\mu^\mp + \ge 
\mbox{3 jets}+ {E\!\!\!/}_T$ comes from 
$e^+ e^- \rightarrow t \bar t g$.

The simulation of the signal and background has been performed
for one of the MSSM representative points chosen for
detailed case studies at the ECFA/DESY Workshop \cite{tesla}: a 
mSUGRA scenario defined by $m_0=100$ GeV, $M_{1/2}=200$ GeV, $A_0=0$
GeV, $\tan\beta=3$ and ${\rm sgn}(\mu)=+$.  A simple parton level
simulation has been performed with a number of kinematic cuts listed
in \cite{GKR}.   
For the processes (\ref{snulfv}) and (\ref{charlfv}) we
find after cuts the following cross-sections,
$\chi_{23}(3-4 \chi_{23}) \sin^2 2\tilde{\theta}_{23} \times 0.51$ fb and
$\chi_{23} \sin^2 2\tilde{\theta}_{23} \times 0.13$ fb,
respectively, while the background amounts to 0.28 fb.

\begin{figure}
\hspace*{2cm} $\Delta \tilde{m}_{23}$/GeV 
\vspace{-1cm}
 \begin{center}
\epsfig{figure=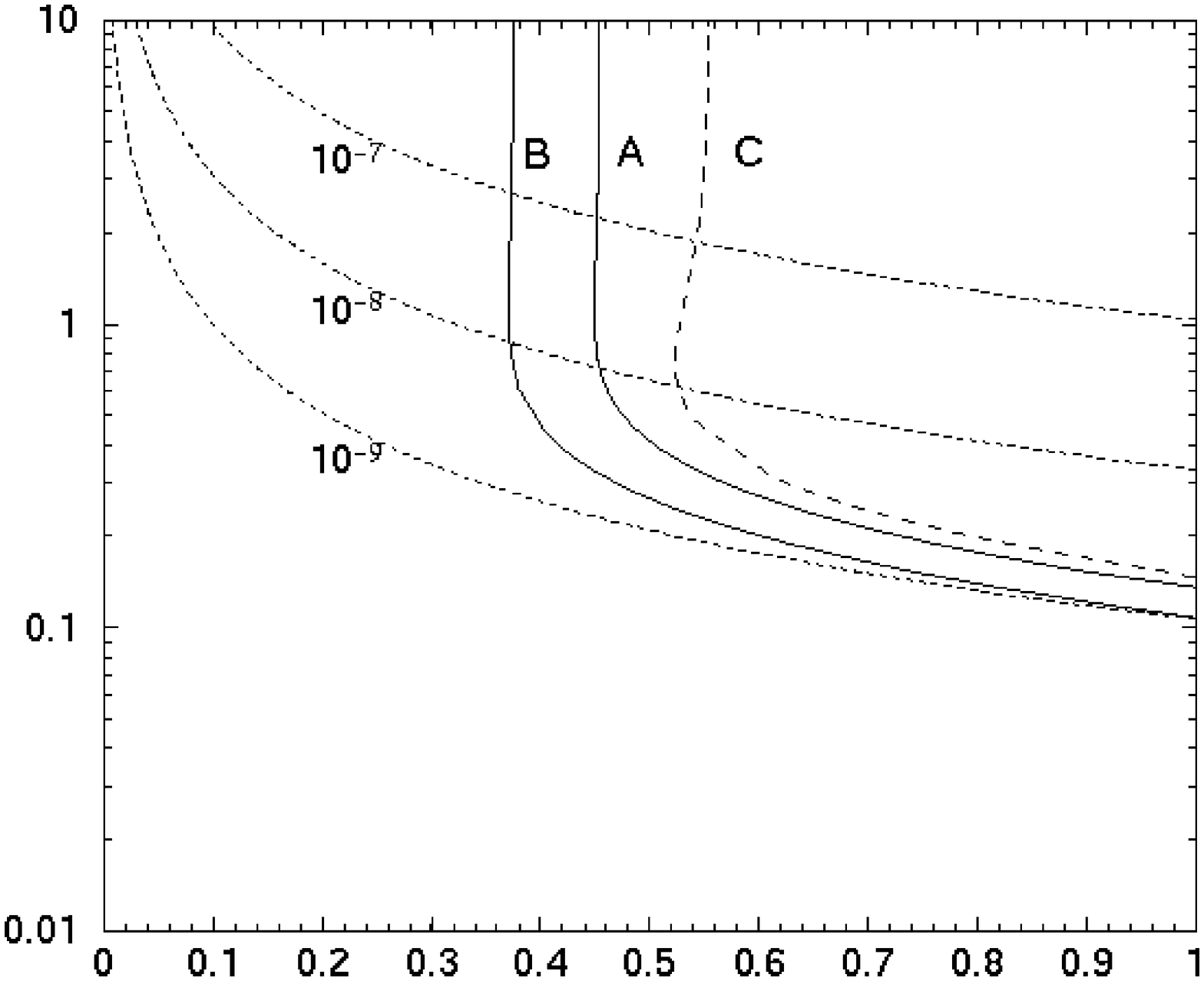,width=8cm,height=7cm}\\
$\sin2\tilde{\theta}_{23}$
 \end{center}
%\vskip -1cm
\caption{
Various 3$\sigma$ significance contours in 
the $\Delta \tilde{m}_{23} -\sin2\tilde{\theta}_{23}$ plane,
for the SUSY point mentioned in the text.
The contours A and B show the integrated signals
(\protect{\ref{snulfv}}--\protect{\ref{charlfv}}) 
at $\sqrt{s}=$500 GeV and for 500 fb$^{-1}$ and 1000 fb$^{-1}$, 
respectively.
The contour C shows the $\tilde\nu \tilde\nu^c$ contribution separately 
for 500 fb$^{-1}$ \cite{GKR}. The dotted lines indicate contours for
$Br(\tau\to\mu\gamma)$=10$^{-7}$, 10$^{-8}$ and 10$^{-9}$ \cite{jk02}.}
\label{fig_jk}
\end{figure}

In Fig.~\ref{fig_jk} the significance is given by $\sigma_d =
\frac{S}{\sqrt{S+B}}$ where S and B are the numbers of signal and
background events, respectively, for a given
luminosity. Shown is the
region (to the right of the curves) in the $\Delta \tilde{m}_{23} -
\sin2\tilde{\theta}_{23}$ plane that can be explored or ruled out at a
3$\sigma$ level at a linear collider of energy 500 GeV for the given
integrated luminosity. The contour A is for 500 fb$^{-1}$ and B
for 1000 fb$^{-1}$. For comparison, the boundary C shows the reach in
the process $\tilde{\nu}_i\tilde{\nu}^c_i$ alone (previously studied
in \cite{feng,nojiri}) using our cuts and assuming a luminosity of 500
fb$^{-1}$. The chargino contribution increases the sensitivity range
to $\sin 2\tilde{\theta}_{23}$ by 10-20\%, while the sensitivity to $\Delta
\tilde{m}_{{23}}$ does not change appreciably.

In the same figure, the contour lines for constant branching ratios of
$\tau\to \mu\gamma$ are shown for comparison \cite{jk02}. 
In the limit of small mass
splitting, $Br(\tau\to\mu\gamma)$ can be calculated in the flavour
basis using the mass
insertion technique \cite{gabbiani}.   
In our 2-3 intergeneration mixing scenario the  
radiative process $\tau\to\mu\gamma$   
constrains the combination of parameters    
\begin{equation}
\delta_{\mu\tau}= \sin2\tilde{\theta}_{23} \Delta
\tilde{m}_{23}/{\tilde m}. 
\end{equation}
The contours in Fig.~\ref{fig_jk} 
have been obtained from the approximate formula
of Ref.\cite{FNS}, normalized to the current experimental limit,
\begin{equation}
Br(\tau\to\mu\gamma)\sim 1.1 \times 10^{-6}
\left(\frac{\delta_{\mu\tau}}{1.4}\right)^2
%,
%(\frac{\delta^{LR}_{\mu\tau}}{8.3\times 10^{-3}})^2]
\left(\frac{100 \mbox{
GeV}}{ {\tilde m}}\right)^4 \label{max}.
\end{equation}
This approximation 
only provides an order of magnitude estimate of the upper limit
for the supersymmetric contribution to the radiative lepton decay.  
The exact result, which is sensitive to the details of mass
spectra and mixings, can in fact be much smaller due to cancellations
among different contributions \cite{Deppisch:2002vz}.  
Fig.~\ref{fig_jk} demonstrates  that information from 
slepton production and decay could be competitive to the radiative lepton
decays.
In particular a LC  
can help to explore the 
small $\Delta \tilde{m}_{23}$ region.
%%% INSERT
It should be stressed, though, that in a given model for lepton flavour 
violation also the correlation with $\mu \to e \gamma$ has to be considered
\cite{Deppisch:2003wt},
which in many cases can yield a more severe bound, as discussed in the 
next section.

%%%%%%%%%%%%%%%%%%%%%%%%%%%MERGE%%%%%%%%%%%%%%%%%%%%%%%%%%%%%%%%%%

\section{Case study for the supersymmetric seesaw model}

As a definite and realistic example for SLFV we consider
the seesaw mechanism in mSU\-GRA models.
In supersymmetric theories with heavy right-handed Majorana neutrinos, 
the seesaw mechanism \cite{seesaw}
can give rise to light neutrino masses at or below 
the sub-eV scale. Furthermore, the massive neutrinos affect the 
renormalization group running of the slepton masses, 
generating flavour off-diagonal terms in the mass matrix.
These in turn lead to SLFV in scattering processes at high energies 
and in rare decays. For illustration
of the potential and complementarity of such SLFV searches we
focus on the LC processes $e^{\pm}e^-\rightarrow l_{\beta}^{\pm} 
l_{\alpha}^- \tilde{\chi}_b^0 \tilde{\chi}_a^0$ 
involving slepton pair production and subsequent decay, and
on the corresponding
radiative decay $l_\alpha \to l_\beta \gamma$. In particular,
in an early ATLAS note \cite{serin}
$\tau \to \mu \gamma$ is estimated to 
be observable at the LHC for a branching ratio of order $10^{-7}$.
However, the limit one can reasonably expect may be an order
of magnitude better \cite{denegri}. 

For our study we use the mSUGRA benchmark scenarios proposed in 
\cite{Battaglia:2001zp} for LC studies, concentrating on those
which predict charged left-handed sleptons that are light enough 
to be pair-produced at the center-of-mass energy $\sqrt{s}=500$~GeV.
Furthermore, we implement the seesaw mechanism assuming degenerate
Majorana masses for the right-handed neutrinos and constrain the
neutrino Yukawa couplings by the measured masses and mixings of
the light neutrinos. 
Further sources of SLFV exist in other models such as
GUTs \cite{guts}. However, no
realistic three generation case study of effects 
for collider processes has been 
performed so far, so that we restrict the discussion to the minimal seesaw 
model, here.

%%%%%%%%%%%%%%%%%%%%%%%%%%%%%%%%%%%%%%%%%%%%%%%%%%%%%%%%%%%%%%%%%%
\subsection{Supersymmetric seesaw mechanism}

If three right-handed neutrino singlet fields $\nu_R$
are added to the MSSM particle content, one has  
the additional terms \cite{Casas:2001sr}
\begin{equation}
W_\nu = -\frac{1}{2}\nu_R^{cT} M \nu_R^c + \nu_R^{cT} Y_\nu L \cdot H_2
\label{suppot4}
\end{equation}
in the superpotential.
Here, \(Y_\nu\) is the matrix of neutrino Yukawa couplings, 
$M$ is the right-handed neutrino Majorana mass matrix, and
$L$ and $H_2$ denote the left-handed 
lepton and hypercharge +1/2 Higgs doublets, respectively. 
At energies much below the mass scale $M_R$
of the right-handed neutrinos, 
$W_{\nu}$ leads to the following
mass matrix for the light neutrinos:
\begin{equation}\label{eqn:SeeSawFormula}
M_\nu = Y_\nu^T M^{-1} Y_\nu (v \sin\beta )^2.
\end{equation}
From that the light neutrino masses $m_1$, $m_2$, $m_3$
are obtained after diagonalization by the unitary MNS matrix \(U\).
The basis is chosen such that the matrices of the charged lepton 
Yukawa couplings and Majorana masses are diagonal, which is always 
possible. 

Furthermore, the heavy neutrino mass eigenstates give rise 
to virtual corrections to the slepton mass matrix 
that are responsible for lepton flavour violating processes.
More specifically, in the mSUGRA models considered, 
the mass matrix of the charged sleptons is given by 
\begin{eqnarray}
 m_{\tilde l}^2=\left(
    \begin{array}{cc}
        m_{\tilde l_L}^2    & (m_{\tilde l_{LR}}^{2})^\dagger \\
        m_{\tilde l_{LR}}^2 & m_{\tilde l_R}^2
    \end{array}
      \right)
\label{mslept}
\end{eqnarray}
with
\begin{eqnarray*}
  (m^2_{\tilde{l}_L})_{ij}     \!\!\!&=&\!\!\! (m_{L}^2)_{ij} 
+ \delta_{ij}\bigg(m_{l_i}^2 
+ m_Z^2 
\cos 2\beta \left(-\frac{1}{2}+\sin^2\theta_W \right)\bigg) 
\label{mlcharged} \\
  (m^2_{\tilde{l}_{R}})_{ij}     
\!\!\!&=&\!\!\! (m_{R}^2)_{ij} 
+ \delta_{ij}(m_{l_i}^2 - m_Z^2 \cos 2\beta 
\sin^2\theta_W) \label{mrcharged} \\
 (m^{2}_{\tilde{l}_{LR}})_{ij} 
\!\!\!&=&\!\!\! A_{ij}v\cos\beta-\delta_{ij}m_{l_i}\mu\tan\beta.
\end{eqnarray*}
When $m^2_{\tilde{l}}$
is evolved from the GUT scale \(M_X\) to the electroweak scale 
characteristic for the experiments, one obtains
\begin{eqnarray}
m_{L}^2\!\!\!&=&\!\!\!m_0^2\mathbf{1} + (\delta m_{L}^2)_{\textrm{\tiny MSSM}} 
+ \delta m_{L}^2 \label{left_handed_SSB} \\
m_{R}^2\!\!\!&=&\!\!\!m_0^2\mathbf{1} + (\delta m_{R}^2)_{\textrm{\tiny MSSM}} 
+ \delta m_{R}^2 \label{right_handed_SSB}\\
A\!\!\!&=&\!\!\!A_0 Y_l+\delta A_{\textrm{\tiny MSSM}}+\delta A \label{A_SSB},
\end{eqnarray}
where $m_{0}$ is the common soft SUSY-breaking scalar mass and $A_{0}$ the 
common trilinear coupling. The terms 
\((\delta m_{L,R}^2)_{\textrm{\tiny MSSM}}\) and 
\(\delta A_{\textrm{\tiny MSSM}}\) are well-known flavour-diagonal MSSM
corrections. In addition, the evolution generates the
off-diagonal terms $\delta m^2_{L,R}$ and $\delta A$
which, in leading-log approximation and for degenerate 
right-handed Majorana masses \(M_i=M_R,i=1,2,3\),
are given by \cite{Hisano:1998fj}
\begin{eqnarray}\label{eq:rnrges}
  \delta m_{L}^2 \!\!\!&=&\!\!\! -\frac{1}{8 \pi^2}(3m_0^2+A_0^2)(Y_\nu^\dag Y_\nu) 
\ln\left(\frac{M_X}{M_R}\right) \label{left_handed_SSB2}\\
  \delta m_{R}^2\!\!\! &=&\!\!\! 0  \\
  \delta A\!\!\! &=&\!\!\! -\frac{3 A_0}{16\pi^2}(Y_l Y_\nu^\dag Y_\nu) 
\ln\left(\frac{M_X}{M_R}\right).
\end{eqnarray}

In order to determine the product $Y_\nu^\dagger Y_\nu$ 
of the neutrino Yukawa coupling matrix entering these corrections, 
one uses the expression   
\begin{eqnarray}\label{eqn:y}
 Y_\nu  = \frac{\sqrt{M_R}}{v\sin\beta} R \cdot 
\textrm{diag}(\sqrt{m_1},\sqrt{m_2},\sqrt{m_3}) \cdot  U^\dagger,
\end{eqnarray}
which follows from \(U^T M_\nu U = \textrm{diag}(m_1,m_2,m_3)\) and 
(\ref{eqn:SeeSawFormula}) \cite{Casas:2001sr}.
Here, \(R\) is an unknown complex orthogonal matrix parametrizing
the ambiguity in the relation of Yukawa coupling and mass matrices. 
In the following we will assume \(R\) to be real which suffices for
the present purpose. In this case, 
\(R\) drops out from the product $Y_\nu^\dagger Y_\nu$,
\begin{eqnarray}\label{eqn:yy}
 Y_\nu^\dagger Y_\nu = \frac{M_R}{v^2\sin^2\beta} U \cdot
\textrm{diag}(m_1,m_2,m_3) \cdot U^\dagger.
\end{eqnarray}
Using existing neutrino data on the mass squared differences  
and the mixing matrix \(U\) together with bounds and assumptions 
on the absolute mass scale
one can calculate 
$Y_\nu^\dagger Y_\nu$. The only free parameter is the Majorana 
mass scale $M_R$. The result is 
then evolved
to the unification scale $M_X$ and
used as an input in the renormalization group corrections 
(\ref{eq:rnrges}) to the slepton mass matrix.
Finally, diagonalization of (\ref{mslept}) yields the 
slepton mass eigenvalues 
\(\tilde m_i\) and eigenstates \(\tilde{l}_i\) (\(i=1,2,...6)\).
  
%%%%%%%%%%%%%%%%%%%%%%%%%%%%%%%%%%%%%%%%%%%%%%%%%%%%%%%%%%%%%%%%%%
\subsection{Lepton flavour violating processes}

The flavour off-diagonal elements (\ref{eq:rnrges})
in  \(m_{\tilde l}^2\) ($\delta A=0$ in the mSUGRA
scenarios of \cite{Battaglia:2001zp})
induce, among other SLFV effects,
the processes $e^+e^- \to \tilde{l}^+_j \tilde{l}_i^-\to 
l^+_{\beta}l_{\alpha}^-\tilde{\chi}^0_b\tilde{\chi}^0_a$,
where SLFV can occur in the production and decay vertices.
The helicity amplitudes for the 
pair production of \(\tilde{l}_j^+\) and \(\tilde{l}_{i}^-\), 
and the corresponding decay amplitudes 
are given explicitly in \cite{Deppisch:2003wt}. 
In the approximation (\ref{crosspair}) for $\sigma^{\rm pair}_{\alpha \beta}$
one finds
\begin{eqnarray}
\sigma^{\rm pair}_{\alpha\beta}\propto \alpha^4
\frac{|(\delta{m}_L)^2_{\alpha\beta}|^2}{
\tilde{m}^2 \Gamma^2} \;
\sigma(\bar f\,f\to \tilde{l}^+_\alpha \,
\tilde{l}^-_\alpha) 
Br(\tilde{l}^+_\alpha \to l^+_\alpha \, \tilde{\chi_0})  
Br(\tilde{l}^-_\alpha \to  l^-_\alpha\, \tilde{\chi_0})
\label{full_M_squared}
\end{eqnarray}
In the numerical evaluation no slepton degeneracy has been assumed
as in (\ref{propags}), and the amplitude 
for the complete \(2 \to 4\) processes
is summed coherently over the intermediate
slepton mass eigenstates. 

Similary, the terms (\ref{eq:rnrges}) are responsible 
for SLFV radiative decays \(l_\alpha\rightarrow l_\beta \gamma\)
induced by photon-penguin type diagrams with 
charginos/sneutrinos or neutralinos/charged sleptons in the loop.
Again schematically, the
decay rates are given by \cite{Casas:2001sr,Hisano:1998fj}
\begin{equation}
\Gamma(l_\alpha \rightarrow l_\beta \gamma) 
\propto \alpha^3 m_{l_\alpha}^5 
\frac{|(\delta m_L)^2_{\alpha \beta}|^2}{\tilde{m}^8} 
\tan^2 \beta,
\end{equation}
where $\tilde m$ stands for the relevant sparticle masses in the loop.

\begin{figure}[t]
\centering
\includegraphics[clip,scale=0.7]{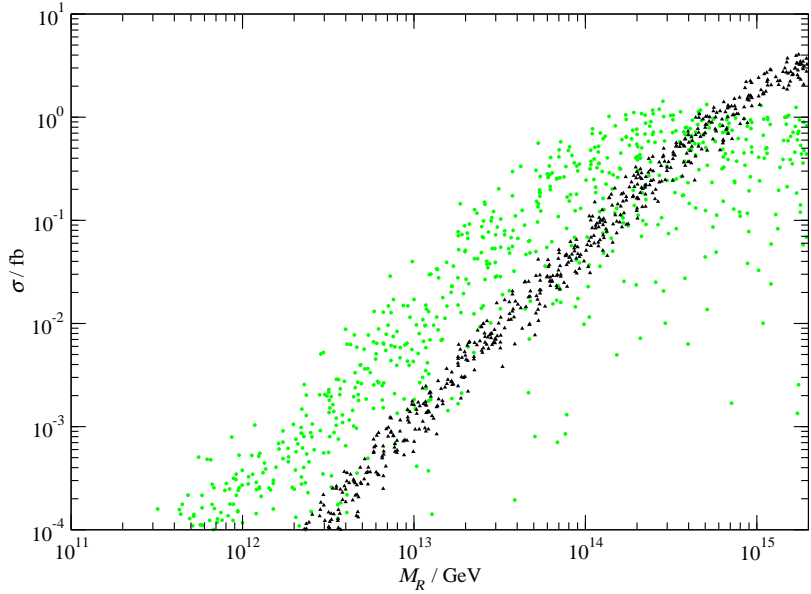}
     \caption{Cross-sections at \(\sqrt{s}=500\) GeV
              for \(e^+e^- \to \mu^+e^- +2\tilde\chi_1^0\) (circles)
              and  \(e^+e^- \to \tau^+\mu^- +  2\tilde\chi_1^0\) (triangles) 
              in scenario B.}
     \label{fig:ep}
\end{figure}

\begin{figure}[t]
\centering
\includegraphics[clip,scale=0.7]{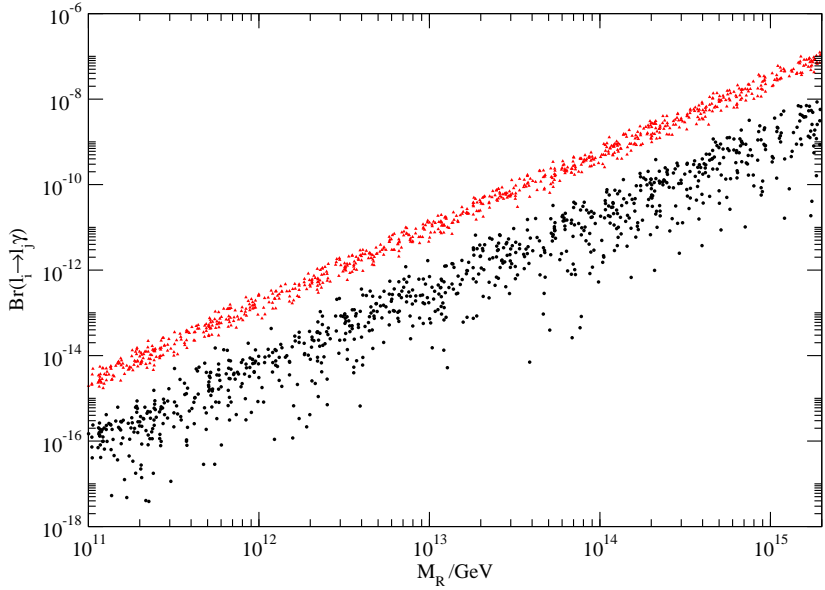}
     \caption{Branching ratios $Br(\tau \to \mu \gamma)$ (upper)
and $Br(\mu \to e \gamma)$ (lower)
in scenario B.}
     \label{fig:mue-taumu}
\end{figure}

\begin{table}[h!]
\begin{center}
\begin{tabular}{|c|c|c|c||c|c|c|}\hline
Scenario & $m_{1/2}$/GeV  & $m_{0}$/GeV & $\tan\beta$ & 
${\tilde m}_{6}$/GeV &${\tilde \Gamma}_6$/GeV & 
$m_{\tilde{\chi}_1^0}$/GeV  \\ \hline\hline
B & 250  & 100  & 10  & 208 & 0.32 & 98  \\ \hline
C & 400  & 90   & 10  & 292 & 0.22 & 164  \\ \hline
G & 375  & 120  & 20  & 292 & 0.41 & 154  \\ \hline
I & 350  & 180  & 35  & 313 & 1.03 & 143  \\ \hline
\end{tabular}
\end{center} 
\caption{\label{mSUGRAscen} Parameters of selected 
mSUGRA benchmark scenarios (from \protect{\cite{Battaglia:2001zp}}). 
The sign of \(\mu\) is chosen to be positive and \(A_0\) is set to 
zero. Given are also the mass and total width of the heaviest
charged slepton and the mass of the lightest neutralino.}
\end{table}

%%%%%%%%%%%%%%%%%%%%%%%%%%%%%%%%%%%%%%%%%%%%%%%%%%%%%%%%%%%%%%
\subsection{Signals and background}

Among the mSUGRA benchmark scenarios proposed in 
\cite{Battaglia:2001zp} for LC studies,
the models B, C, G, and 
I (see Tab.\ref{mSUGRAscen}) 
predict left-handed sleptons which can be pair-produced 
at \(e^+e^-\) colliders \(\sqrt{s}=500\div800\)~GeV cms energies. 
In the following we will confine ourselves 
to these models.

Most likely, at the time when a linear collider will 
be in operation, more precise measurements of the
neutrino parameters will be available than today.
In order to simulate the expected improvement, 
we take the central values of the mass squared 
differences $\Delta m^2_{ij}=|m_i^2-m_j^2|$
and mixing angles $\theta_{ij}$ from a global fit to existing 
data \cite{Gonzalez-Garcia:2000sq} with errors that indicate 
the anticipated 90 \% C.L. intervals of running and proposed 
experiments as further explained in \cite{Deppisch:2002vz}: 
\begin{eqnarray}
&&\tan^2\theta_{23}=1.40^{+1.37}_{-0.66},~~
\tan^2\theta_{13}=0.005^{+0.001}_{-0.005},~~
\tan^2\theta_{12}=0.36^{+0.35}_{-0.16}, \label{nupar1}\\
&&\Delta m_{12}^2=3.30^{+0.3}_{-0.3}\cdot 10^{-5}\textrm{ eV}^2 ,~~
\Delta m_{23}^2=3.10^{+1.0}_{-1.0}\cdot 10^{-3}\textrm{ eV}^2.
\end{eqnarray}
Furthermore, for the lightest neutrino we assume the mass range
\(m_1 \approx 0-0.03\)~eV, which at the lower end corresponds to
the case of a hierarchical spectrum. Towards the upper end, it approaches
the degenerate case.

\begin{figure}[t]
\centering
\includegraphics[clip,scale=0.7]{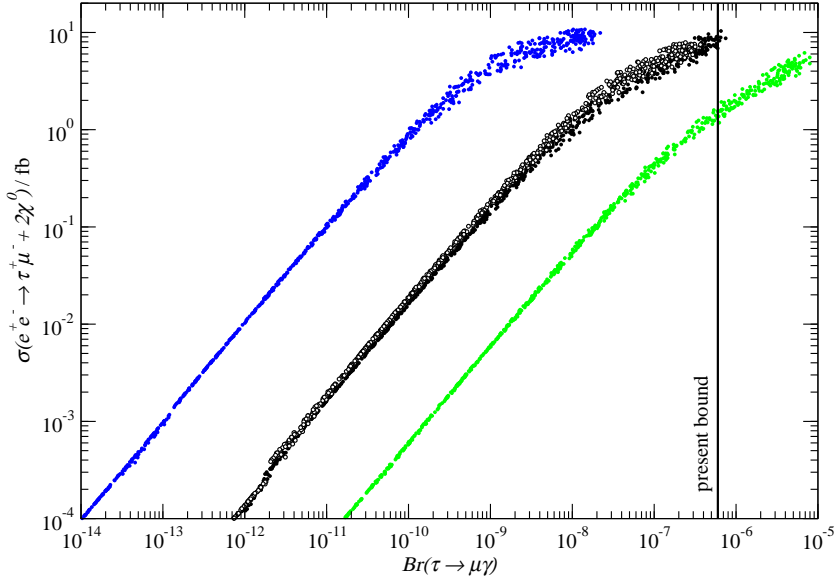}
     \caption{Correlation of \(\sigma(e^+e^- \to \tau^+\mu^- +2\tilde\chi_1^0)\)
              at \(\sqrt{s}=800\) GeV  
              with \(Br(\tau\to \mu\gamma)\) in scenario (from left to right) 
              C, G (open circles), B and I.}
     \label{fig:mutau_lowhigh}
\end{figure}

\begin{figure}[t]
\centering
\includegraphics[clip,scale=0.7]{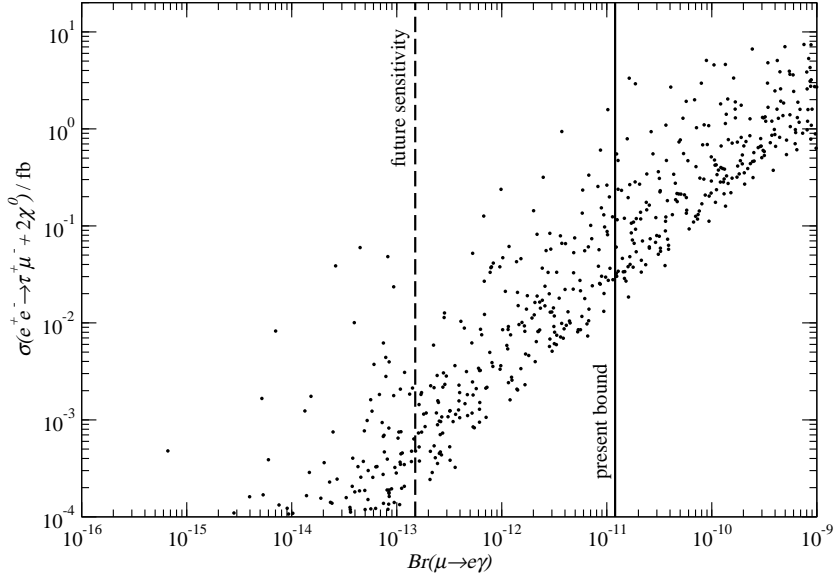}
     \caption{
Correlation of \(\sigma(e^+e^- \to \tau^+\mu^- +2\tilde\chi_1^0)\)
              at \(\sqrt{s}=800\) GeV  
              with \(Br(\mu\to e\gamma)\) in scenario B.}
     \label{fig:mutau800}
\end{figure}

%%%%%%%%%%%%%%%%%%%%%%%%%%%%%%%%%%%%%%%%%%%%%%%%%%%%%%%%%%%%%%%%

In Fig.~\ref{fig:ep}, the cross-sections for 
\(e^+e^-\rightarrow \mu^+ e^- +2\tilde{\chi}^0_1\) and 
\(e^+e^-\rightarrow \tau^+ \mu^- +2\tilde{\chi}^0_1\) 
are plotted for model B. 
The channel \(\tau^+ e^- +2\tilde{\chi}_1^0\) is not
shown since it is strongly suppressed by the small mixing angle 
\(\theta_{13}\), and therefore more difficult to observe.
As can be seen, for a sufficiently large Majorana mass scale 
the SLFV cross-sections can reach several fb. 
The spread of the predictions reflects the uncertainties in the neutrino 
data.

%%%%%%%%%%%Background%%%%%%%%%%%%%%%%%%%%%%%%%%%%%%%%%%%%%%%%%%%%%%%%%%%%%%%%%
The Standard Model background mainly comes from $W$-pair production, 
$W$ production with $t$-channel photon exchange, and $\tau$-pair production.
A 10 degree beam pipe cut and   
cuts on the lepton energy and missing energy reduce the SM background 
cross-sections to less than 30 ~fb for \((\mu e)\) final states and less
than 10 ~fb for \((\tau \mu)\) final states. If one requires a signal
to background ratio, $S/\sqrt{S+ B} = 3$,
and assumes a typical signal cross-section
of 0.1~fb, one can afford a background of about 1~fb. 
Here an integrated luminosity of 1000~fb$^{-1}$ has been assumed.
Whether or not the background process estimate above 
can be further suppressed to this level by applying  
selectron selection cuts, for example, on the acoplanarity, 
lepton polar angle and missing transverse momentum has to be studied in 
dedicated simulations. 
For lepton flavour conserving processes it has been shown
that the SM background 
to slepton pair production can be reduced to about 2-3~fb
at $\sqrt{s}=500$~GeV \cite{Becker:1993fw}.
 
The MSSM background is dominated by
chargino/slepton production with
a total cross-section of
0.2-5~fb and 2-7~fb for \((\mu e )\) 
and \((\tau \mu)\) final states, 
respectively, depending on the SUSY scenario
and the collider energy.
The MSSM
background in the ($\tau e$) channel can also
contribute to the \(\mu e\)
channel via the decay $\tau \rightarrow \mu \nu_{\mu} {\nu}_{\tau}$.
If $\tilde{\tau}_1$ and $\tilde{\chi}^+_1$ are very light, like in scenarios B 
and I, this background can be as large as 20~fb.
However, such events typically contain two neutrinos in addition to the
two LSPs which are also present in the signal events. Thus, after $\tau$ decay 
one has altogether
six invisible particles instead of two, which may allow to discriminate 
the signal 
in $\mu^+e^-+\;/\!\!\!\!E$ also from this potentially 
dangerous MSSM background 
by cutting on various distributions.
But also here one needs a dedicated simulation study, in order to make more 
definite statements.

The corresponding branching ratios,
$Br(\mu \to e \gamma)$ and $Br(\tau \to \mu \gamma)$, 
in model B are displayed in 
Fig.~\ref{fig:mue-taumu}
\cite{Deppisch:2002vz}. 
One sees that
a positive signal for $\mu \to e \gamma$ at 
the minimum branching 
ratio observable in the new PSI experiment,
$Br(\mu \to e \gamma)\simeq 10^{-13}$ \cite{PSI}
would imply a value of 
$M_R$ between $2\cdot 10^{12}$~GeV and $2\cdot 10^{13}$~GeV.
In comparison to $\mu \to e \gamma$ the channel
\(\tau\to\mu\gamma\)
is less affected by the neutrino uncertainties. 
If the sensitivity goal \(Br(\tau\to\mu\gamma)=10^{-8}\) \cite{denegri}
at the LHC is reached
one could probe $M_R=10^{15}$~GeV.

Particularly
interesting and useful are the 
correlations between SLFV in radiative decays and 
slepton pair production. Such a correlation is illustrated in 
Fig.~\ref{fig:mutau_lowhigh} for 
\(e^+e^-\rightarrow \tau^+\mu^- +2\tilde{\chi}_1^0\)
and \(Br(\tau\to \mu \gamma)\).
One sees that the neutrino uncertainties 
drop out, 
while the sensitivity to the mSUGRA parameters remains.
An observation of $\tau \to \mu \gamma$ with the branching ratio
$10^{-8}$ at the LHC would be compatible   
with a cross-section of order 10 fb for 
$e^+e^- \to \sum_{i,j}\tilde{l}_j^+\tilde{l}^-_i\to 
\tau^+ \mu^- +2\tilde{\chi}^0_1$, at least in model C. 
However, there are also correlations of different flavor channels.
This is illustrated 
in Fig.~\ref{fig:mutau800}, where the correlation of 
\(e^+e^-\rightarrow \tau^+\mu^- +2\tilde{\chi}_1^0\)
and \(\mu\to e \gamma\) is shown. Despite of the uncertainties 
from the neutrino sector, 
already the present experimental bound 
\(Br(\mu\to e \gamma)<1.2 \cdot 10^{-11}\)
yields a stronger constraint on $\sigma(e^+e^- \to
\tau^+ \mu^- +2 \tilde\chi_1^0)$
than the one obtained
from Fig.~\ref{fig:mutau_lowhigh}, making 
cross-sections larger than a few $10^{-1}$~fb  
at \(\sqrt{s}=800\)~GeV very unlikely in model B.
If this scenario is correct,
non-observation of 
$\mu \rightarrow e \gamma$
at the new PSI experiment will exclude
the observability of this channel at a LC.
As a final remark we stress that in the channel $e^+e^- \to
\mu^+ e^- +2 \tilde\chi_1^0$ cross-sections of
1~fb are compatible with the present bounds, while no signal at the 
future PSI sensitivity would constrain this channel to less than 0.1~fb. 
However we want to emphasize again that these statements are very model 
dependent, and much bigger cross-sections are possible in general, as shown 
in section \ref{sfcs}.

%%%%%%%%%%%%%%%%%%%%%%%%%%%%%%%%%%%%%%%%%%%%%%%%%%%%%%%%%%%%%%

\section{Summary and outlook}
If superpartners are discovered at future colliders, 
we advocate the search for SUSY lepton flavour violation 
as a high priority topic of the experimental programme. 
At a LC, the most favourable signals are expected to come from
the production and decay of sleptons and charginos. 
Considering only LFV in the $\mu-\tau$ sector, 
a case motivated by the large atmospheric neutrino mixing 
but more difficult to detect than LFV in the $e-\mu$ sector 
due to the presence of decaying taus,
we have shown that the LC measurements may be complementary to 
searches for the radiative $\tau$ decay at the LHC. 
For example, a measurement  of 
$Br(\tau \to \mu \gamma) = 10^{-8}$ at the LHC  
combined with the SLFV signal at a LC would point to 
$\sin2\tilde\theta_{23}\geq 0.4$ and $\Delta\tilde m_{23}\simeq 0.3 - 1$ 
GeV. 
 
In the context of the SUSY seesaw
mechanism of neutrino mass generation,  
correlations between SLFV in radiative decays and 
slepton pair production have been found particularly
interesting. For instance, in a given MSSM scenario the
measurement of $\tau \rightarrow \mu \gamma$ at the LHC
would imply a definite cross section for  
$e^+e^- \to \tau^+ \mu^- +2 \tilde\chi_1^0$ at the LC.
Assuming a reasonable set of MSSM benchmark scenarios and 
$Br(\tau \to \mu \gamma) = 10^{-8}$ and using the 
present neutrino data, one predicts
$\sigma(e^+e^- \to \tau^+ \mu^- +2 \tilde\chi_1^0)$
in the range 0.05 to 10 fb. However, the
non-observation of $\mu \to e \gamma$ with
a branching ratio of about $10^{-13}$ 
at the new PSI experiment would exclude
the observability of $\sigma(e^+e^- \to
\tau^+ \mu^- +2 \tilde\chi_1^0)$ at a linear collider.
While the former correlation involving the same lepton flavours 
is insensitive to the uncertainties in the neutrino 
data (Fig. \ref{fig:mutau_lowhigh}), 
the latter correlation is somewhat smeared out 
(Fig. \ref{fig:mutau800}). 
However, both types of correlations remain sensitive to the mSUGRA 
parameters and, hence, provide very useful
tools for probing the origin of lepton flavour violation.
 
The complementarity of the LHC and LC (and of low-energy experiments) 
in the context of lepton
flavour violation is far from being exhausted by the present study. 
Quantitative analyses of
the impact of precise mass measurements at the LC on identifying the LFV
decay chains at the LHC (and vice-versa) and other important features  
call for detailed Monte Carlo simulations
which should be undertaken in the next round of the LHC/LC studies.

%%%%%%%%%%%%%%%%%%%%%%%%%%%%%%%%%%%%%%%%%%%%%%%%%%%%%%%%%%%%%%

\subsection*{Acknowledgements}

This work was performed in the framework of the LHC/LC study group
and supported by the Bundesministerium f\"ur Bildung und 
Forschung (BMBF, Bonn, Germany) under 
the contract number 05HT4WWA2.
The work of JK was supported by the KBN Grant 2 P03B 040 24
(2003-2005) and 115/E-343/SPB/DESY/P-03/DWM517/2003-2005.

\end{document}